# Non-Universal Critical Behaviors in Disordered Pseudospin-1 Systems

A. Fang[1], Z. Q. Zhang[1], Steven G. Louie[2,3,4] and C. T. Chan[1,*]

[1]Department of Physics, The Hong Kong University of Science and Technology, Clear Water Bay, Hong Kong, China

[2]Institute for Advanced Study, The Hong Kong University of Science and Technology, Clear Water Bay, Hong Kong, China

[3]Department of Physics, University of California at Berkeley, Berkeley, CA 94720, USA

[4]Materials Sciences Division, Lawrence Berkeley National Laboratory, Berkeley, CA 94720, USA

*Email: phchan@ust.hk

## Abstract

It is well known that for ordinary one-dimensional (1D) disordered systems, the Anderson localization length $\xi$ diverges as $\lambda^m$ in the long wavelength limit ($\lambda \to \infty$) with a universal exponent $m=2$, independent of the type of disorder. Here, we show rigorously that pseudospin-1 systems exhibit non-universal critical behaviors when they are subjected to 1D random potentials. In such systems, we find that $\xi \propto \lambda^m$ with $m$ depending on the type of disorder. For binary disorder, $m=6$ and the fast divergence is due to a super-Klein-tunneling effect (SKTE). When we add additional potential fluctuations to the binary disorder, the critical exponent $m$ crosses over from 6 to 4 as the wavelength increases. Moreover, for disordered superlattices, in which the random potential layers are separated by layers of background medium, the exponent $m$ is further reduced to 2 due to the multiple reflections inside the background layer. To obtain the above results, we developed a new analytic method based on a stack recursion equation. Our analytical results are in excellent agreements with the numerical results obtained by the transfer-matrix method (TMM). For pseudospin-1/2 systems, we find both numerically and analytically that $\xi \propto \lambda^2$ for all types of disorder, same as ordinary 1D disordered systems. Our new analytical method provides a convenient way to obtain easily the critical exponent $m$ for general 1D Anderson localization problems.

# I. INTRODUCTION

Realistic systems are never perfectly ordered, and hence studying the effects of disorder is essential in understanding the transport behavior in a vast variety of electronic and classical wave materials [1-30]. Among all disorder-induced phenomena, Anderson localization [1-6] is perhaps the most fundamental and universal. As a wave localization behavior stemming from the wave interference effect, prior works show that Anderson localization exhibits universal behaviors which are independent of the type of disorder and the details of the random potential [6-10]. For example, for ordinary 1D disordered systems, the Anderson localization length $\xi$ diverges as $\lambda^m$ with a universal exponent $m = 2$ [6-10] in the long wavelength limit ($\lambda \to \infty$), independent of the type of disorder unless the disorder is correlated in some special ways [6, 16-19]. The effect due to random potential distributions enters only as a prefactor in the asymptotic behavior of the localization length, i.e., $\xi \propto \lambda^2 / <\delta V^2>_c$ (see for example Refs.5-10, and section 2.1 of Supplemental Material [31]), where $<>_c$ denotes ensemble averaging and $<\delta V^2>_c$ is the second moment of the random potential distribution. In this work, we show that the commonly accepted universal critical behavior does not hold for pseudospin-1 disordered systems. For pseudospin-1 systems subjected to 1D random potentials, we discovered that the details of the random potential can affect the exponent $m$ directly, i.e., the value of $m$ depends strongly on the type of disorder. To the best of our knowledge, we have not seen any other disordered systems which have such non-universal critical behavior.

Pseudospin systems can be realized using two-dimensional (2D) materials exhibiting conical band dispersions, with two or three bands intersecting linearly at a point, usually referred as the Dirac or Dirac-like point [32-53]. The physics near the nodal point can be described by an effective spin-orbit interaction. The 1D random potential mentioned above refers to the case where the potential only fluctuates along one specific direction, although the pseudospin systems are in 2D. The best known example of pseudospin materials is graphene [32-34], with a pseudospin $S = 1/2$. Its low energy excitations can be described by a massless Dirac equation, and the orbital wave function can be represented by a two-component spinor, with each component corresponding to the amplitude of the electron wave function on one of the trigonal sublattices of graphene. We emphasize that the pseudospin here is not the intrinsic spin of electrons, but refers to the spatial degrees of freedom. The Dirac cone and the associated pseudospin-1/2 characteristic of quasiparticles can also be found in other systems such as topological insulators [35, 36] and the photonic and phononic counterparts of graphene [37, 38].

Another interesting example is pseudospin-1 material which possesses a threefold degeneracy at a Dirac-like point, where two cones meet and intersect with an additional flat band [39-53]. Such threefold degeneracy can be realized in some 2D dielectric photonic crystals (PCs) in which the accidental degeneracy of monopole and dipole excitations gives three degrees of freedom [39-41]. The photon transport in such PCs is governed by an effective spin-orbit Hamiltonian with pseudospin $S = 1$ and the wave functions are described by a three-component spinor [41]. Materials with such Dirac-like cones have also been experimentally realized in ultracold atom [42], photonic [43-45] and electronic [46, 47] Lieb lattices. Moreover, prior theoretical works predict that the Dirac-like cone can be found in artificial crystals of ultracold atoms with a Dice (or $T_3$) lattice [48-50] and certain electronic materials such as blue phosphorene oxide [52] and $SrCu_2(BO_3)_2$ [53]. In graphene, electron transport can be controlled by imposing a gate voltage to shift rigidly the conical dispersion. In analogy with the gate voltage in graphene, the potential shifts of pseudospin-1 PCs and ultracold atom systems can be emulated by a change of length scale [41] and an appropriate holographic mask [48-50], respectively.

Due to the conical band structure and the chiral nature of the underlying quasiparticle states, both pseudospin-1/2 and -1 systems share some common transport properties, such as one-way transport [34, 41, 51] and the supercollimation of a wave packet in a superlattice [25, 54, 55]. For both pseudospin -1/2 and -1 systems subjected to 1D random potentials, waves propagating in the direction normal to the fluctuating potential barriers are delocalized due to the one-way transport phenomenon [23, 24, 56, 57]. For obliquely incident waves, there exists a minimum localization length at some critical disorder strength for both systems and additional disorder makes the waves less localized [24]. However, different pseudospin number also gives rise to distinct physical behaviors. For example, for pseudospin-1/2 systems, an eigenmode trajectory encircling the Dirac point picks up a Berry phase of $\pi$, which in turn gives rise to a topological delocalization effect for systems subjected to 2D disordered potentials [29, 30]. Such delocalization effect does not occur in pseudospin-1 systems due to a zero Berry phase [41]. In addition, in the presence of 1D potential barrier, the so-called super-Klein tunneling effect (SKTE), which is the perfect transmission for all incident angles when the incident energy equals half of the barrier, can exist only in pseudospin-1 systems [41, 51].

Here, we report a new non-universal localization behavior which is unique to pseudospin-1 systems. We find that for pseudospin-1 systems subjected to 1D random potentials, the Anderson localization length $\xi$ at a fixed incident angle diverges as $\lambda^m$, with the exponent $m$ depending on the type of disorder. For the case of binary disorder, the Anderson localization length $\xi$ diverges as $\lambda^6$ in the long wavelength limit due to the SKTE. If we add additional randomness to the binary disorder, the SKTE breaks down and the divergence of $\xi$ crosses over from $\lambda^6$ to $\lambda^4$ as $\lambda$ increases. Furthermore, for disordered superlattices, $\xi$ diverges at an even lower rate of $\lambda^2$ due to the multiple reflections in layers of background medium. We discovered the new Anderson localization behaviors using an analytic stack recursion equation with some proper choice of scattering elements, from which exact asymptotic solutions of $\xi$ can be derived in the long wavelength limit for different types of disorder. These solutions are quantitatively reproduced and confirmed by our numerical simulations using the transfer-matrix method (TMM).

We have also studied the long-wavelength behaviors of Anderson localization length for pseudospin-1/2 systems subjected to 1D random potentials. Similar to the case of ordinary materials, we find both analytically and numerically that a universal $\lambda^2$ behavior exists in all types of disorder. It should be stressed that the analytic method proposed here provides a simple way to obtain quickly the critical exponent $m$ through the long-wavelength scattering properties of individual scattering elements. Our method is applicable to other 1D Anderson localization problems as long as the system considered possesses a divergent localization length in the long wavelength limit.

Although the non-universal 1D Anderson localization behaviors exhibited in pseudospin-1 systems cannot be found in ordinary 1D disordered systems, they still have some transport properties in common. It is well known that the Anderson localization length in 1D is of the same order as the system's transport mean free path [5, 6] as can be derived from the self-consistent theory of Anderson localization [5, 58, 59]. This result implies the absence of a diffusive regime in 1D. To check this point, we have studied numerically the transport mean free path $l_t$ for both pseudospin-1 and pseudospin-1/2 systems for the above three types of disorder. We find that the transport mean free path does diverge as $\lambda^m$ with the same critical exponent $m$ as that of $\xi$ for both pseudospin systems and each type of disorder. This implies the absence of a diffusive regime in pseudospin systems, same as the case of ordinary 1D disordered systems.

## II. DISORDERED PSEUDOSPIN-1 MODEL

The system under investigation consists of $N$ layers of pseudospin-1 systems stacked together and subjected to 1D random potentials along the stacking direction (see Fig. 1). Each layer will be described using a pseudospin-1 Hamiltonian and it has equal thickness $d$, and the potential in the $i$-th random layer is $v_i$. Here we take the Dirac-like point of the background medium as the origin of energy, i.e., $V = 0$. We consider a plane wave impinging on the layered structure from the background medium at an incident angle $\theta$ with incident energy $E$. We only consider oblique incidence ($\theta \neq 0$) since the normally incident waves ($\theta = 0$) are delocalized in a 1D random potential due to the one-way transport [23, 24]. The propagation of pseudospin-1 waves in a 1D potential $V(x)$ is governed by the following equation [24, 41, 51]:

$$H\psi = \left[ \hbar v_g \vec{S} \cdot \vec{k} + V(x)I \right]\psi = E\psi \ . \tag{1}$$

Here $\psi = (\psi_1, \psi_2, \psi_3)^T$ is a three-component spinor function, $\vec{k} = (k_x, k_y)$ is the wavevector operator with $k_x = -i\dfrac{\partial}{\partial x}$ and $k_y = -i\dfrac{\partial}{\partial y}$, $\vec{S} = (S_x, S_y)$ is the matrix representation of the spin-1 operator, $v_g$ is the group velocity, and $I$ is a $3 \times 3$ identity matrix. In the case of 1D random potential $V(x)$, the wavevector component parallel to the interface, $k_y$, is a conserved quantity. We note that Eq. (1) holds for both matter waves (e.g., electrons [46, 47, 52, 53] and ultracold atoms [42, 48-50]) and electromagnetic waves [39-41, 43-45] as long as the band dispersion of the system near some high-symmetry point in the Brillouin zone can be described by Eq. (1) with the wavevector $\vec{k}$ measured from the symmetry point. For simplicity, we use normalized energy $\bar{E} = E / \hbar v_g$ and normalized potential $\bar{V}(x) = V(x) / \hbar v_g$ in the following.

To highlight the dependence of the localization critical exponent $m$ on the type of disorder, we consider three types of 1D random potential [see Figs. 1(b)-(d)] commonly studied for Anderson localization [6, 19-22, 26-28]: (I) binary disorder, (II) binary disorder with additional randomness, (III) disordered superlattices. We shall refer to them as Type I, II and III disorder. For Type I disorder, the normalized potential in the $i$-th random layer $\bar{v}_i = v_i / \hbar v_g$ is an independent random variable, which is taken as $\bar{W}$ with probability $p$ and $-\bar{W}$ with probability $1-p$, where $\bar{W} = W / \hbar v_g$ is the strength of binary disorder as shown in Fig. 1(b). For Type II disorder, we take $\bar{v}_i$ as $\bar{W}(1+\delta_i)$ with probability $p$ and $-\bar{W}(1+\delta_i)$ with probability $1-p$ [see Fig. 1(c)]. Here $\delta_i$ is another random number distributed uniformly on an interval $[-Q, Q]$ ($Q < 1$). For Type III disorder, the random potential layers are separated by layers of background medium ($\bar{V} = 0$) and $\bar{v}_i$ takes the form $\bar{v}_i = \bar{U}_0(1+\delta_i)$, where $\bar{U}_0$ is a normalized potential and $\delta_i$ is the same random number adopted for Type II disorder [see Fig. 1(d)]. When $\delta_i = 0$, the system represents a superlattice [26-28, 55]. The Anderson localization length $\xi$, or the inverse of the Lyapunov exponent $\gamma$, can be obtained through the following relation:

$$\xi = \gamma^{-1} = -\lim_{N \to \infty} \frac{2L_N}{\langle \ln T_N \rangle_c}, \tag{1}$$

where $L_N$ is the sample thickness and $T_N$ is the transmission coefficient.

# III. ANDERSON LOCALIZATION LENGTH CALCULATED BY THE TRANSFER-MATRIX METHOD

## A. Pseudospin-1 systems

We first use the TMM [41] to study the Anderson localization length as a function of wavelength $\lambda = 2\pi / \bar{E}$ for three types of 1D random potential described above. Results of averaging over 4000 realizations are shown in Fig. 2 for two incident angles, $\sin\theta = 0.2$ (solid circles) and $\sin\theta = 0.5$ (open circles), and $\bar{W} = 0.2\pi / d$. $N$ is chosen to be five times the localization length. For the case of binary disorder with $p = 0.5$ (Type I disorder), as shown in Fig. 2(a), the log-log plot of $\xi$ vs. $\lambda$ shows a straight line with a slope of 6 at long wavelengths for both incident angles, indicating that $\xi \propto \lambda^6$. In sharp contrast to the $\lambda^2$ dependence found in ordinary 1D disordered materials, the localization length here diverges much more rapidly. The fast divergence is partly due to the so-called SKTE unique to pseudospin-1 systems [41, 51]. For binary disorder, the potential is either $\bar{W}$ or $-\bar{W}$, and $\bar{E} = 0$ ($\lambda \to \infty$) is exactly the midpoint of the potential difference at which the SKTE occurs where transmission is unity for all incident angles [41, 51]. To confirm the role of SKTE in the $\lambda^6$ anomaly, we add additional randomness with a uniform distribution to the binary disorder to destroy the SKTE. For this Type II disorder, the results of $p = 0.5$ and $Q = 0.3$ are plotted in Fig. 2(b). We see that $\xi$ follows closely the behaviors shown in Fig. 2(a) when $20d < \lambda < 50d$, and crosses over to a slower $\xi \propto \lambda^4$ divergence when $\lambda > 200d$. The slowdown from $\lambda^6$ to $\lambda^4$ indicates that the SKTE contributes a $\lambda^2$ factor to the asymptotic exponent. To elaborate on this point, we apply a disorder bias $\bar{V}_0$ to the binary disorder so that the random potential becomes either $\bar{W} + \bar{V}_0$ or $-\bar{W} + \bar{V}_0$. The SKTE now occurs at $\bar{E} = \bar{V}_0$ instead of $\bar{E} = 0$. In Fig. 2(c), we plot the results of $p = 0.5$ and $\bar{V}_0 = 0.1\pi / d$, and find that $\xi$ indeed diverges as $\lambda^4$ at long wavelengths. However, there appears another divergence at $\bar{E} = \bar{V}_0$ (or $\lambda = 2\pi / \bar{E} = 20d$), independent of the incident angle. To extract the exponent of this divergence, we plot $\xi$ vs. $|\bar{E} - \bar{V}_0|$ in log-log scale in Fig. 3, which clearly shows a relation $\xi \propto (\bar{E} - \bar{V}_0)^{-2}$ as $\bar{E} \to \bar{V}_0$. When $\bar{V}_0 \to 0$, this relation contributes an extra $\bar{E}^{-2}$ ($\lambda^2$) factor to the existing $\bar{E}^{-4}$ ($\lambda^4$) behavior in the long wavelength limit, leading to the $\bar{E}^{-6}$ ($\lambda^6$) behavior found in Type I disorder [Fig. 2(a)]. The same localization characteristics are found in binary disorder systems with other values of $p$. The results of $p = 0.3$ are plotted by solid circles in Figs. S2(a)-(c) (see Supplemental Material [31]). For Type III disorder, in which small potential fluctuations are added to a superlattice, we plot the results of $\bar{U}_0 = 0.2\pi / d$ and $Q = 0.3$ in Fig. 2(d). Interestingly, we find that $\xi$ diverges at an even lower rate of $\xi \propto \lambda^2$.

Thus, we find that the critical exponent $m$ of Anderson localization depends strongly on the type of disorder for pseudospin-1 systems subjected to 1D random potentials. To the best of our knowledge, such non-universal critical behavior has not been seen in any ordinary 1D disordered systems, in which the Anderson localization length always diverges as $\lambda^2$ for any disorder without inter-layer correlations [6-10]. It should be pointed out that in the layered media considered here, although the layer thickness $d$ can be considered as the correlation length, such a short-range correlation inside one layer is trivial as it only sets the unit of the localization length. However, certain non-trivial short- and long-range correlations in the random potentials can significantly change the wave localization behavior of a disordered system [6, 16-19, 58, 59]. For example, it was shown that certain short-range correlations between the site energies and hopping elements in a disordered tight-binding model can make a localized wave super-diffusive [19, 59]. Long-range correlations, which decay in a power law manner, can result in the emergence of effective mobility edges even in 1D [19, 58]. In the long wavelength limit, it has been shown that the critical exponent $m$ in 1D depends strongly on specific correlations [16-19]. It should also be pointed out that unlike the situations in the Anderson localization experiments with ultracold atoms [19, 60, 61], in which the particles can be trapped in the few deepest potential wells in the low energy limit, $E = 0$ in our work refers to the Dirac-like point of the background medium and the particles with energies near that point can tunnel through any potential barriers due to the chiral nature of the conical dispersion.

It is well known in ordinary 1D disordered systems that there is no diffusive regime in such systems as

the transport mean free path is of the same order as the Anderson localization length [5, 6, 62, 63]. It is

interesting to know if this is also true for pseudospin-1 systems. To check this point, we calculated the

transport mean free path $l_t$ for the above three types of disorder. Here the transport mean free path $l_t$

( $= N_t d$ ) is defined as the sample thickness at which the ensemble averages of the transmission and reflection coefficients of the sample are equal, i.e., $< T_{N_t = l_t / d} >_c = < R_{N_t = l_t / d} >_c = \frac{1}{2}$ . The numerical results obtained by the TMM are plotted in Figs. 4(a)-(c) for Types I-III disorder, respectively, for two incident angles, $\sin \theta = 0.2$ (solid circles) and $\sin \theta = 0.5$ (open circles). The least mean square fittings (red dotted lines) show that for each type of disorder, the transport mean free path $l_t$ has the similar long-wavelength characteristic as the Anderson localization length $\xi$, i.e., $l_t \propto \lambda^m$ with the same critical exponent $m$ as that for $\xi$ . Furthermore, by comparing the values of $\xi$ and $l_t$ shown in

Figs. 2 and 4, respectively, we find that the magnitude of $\xi$ is about 2~3 times of $l_t$, which is consistent with the known results reported in ordinary 1D disordered systems [5, 6, 62, 63]. This result also implies the absence of a diffusive regime in pseudospin-1 systems, irrespective of the type of disorder.

## B. Pseudospin-1/2 systems

For comparison, we have also studied numerically the localization behaviors for pseudospin-1/2

systems. The Hamiltonian of pseudospin-1/2 systems takes the same form as Eq. (1) except that the wave function $\psi$ is a two-component spinor [25, 34], $\psi = (\psi_1, \psi_2)^T$ and the spin matrices become

Pauli matrices. However, different pseudospin number leads to different boundary conditions between two neighboring layers [41, 51]. For pseudospin-1 systems, $\psi_2$ and $\psi_1 + \psi_3$ are continuous at the

boundary, while for pseudospin-1/2 systems, $\psi_1$ and $\psi_2$ are continuous. We note that even though

the wave function is a three-component spinor for pseudospin-1 systems, there are only two

independent boundary conditions for both matter waves and EM waves [41, 51]. The different

boundary conditions in turn result in different transfer matrices and affect the localization behaviors. In

Figs. 5(a)-(d), we plot the TMM results of localization length as a function of wavelength at two

incident angles for different types of disorder and find the same $\xi \propto \lambda^2$ behavior. In Figs. 6(a)-(c),

the TMM results of the transport mean free path $l_t$ are plotted as a function of wavelength for Types

I-III disorder, respectively. It is found that $l_t \propto \lambda^2$ for all types of disorder and $\xi \approx 2l_t \sim 3l_t$,

indicating that a diffusive regime is also absent in pseudospin-1/2 systems.

# IV. EXACT ASYMPTOTIC SOLUTIONS OF ANDERSON LOCALIZATION LENGTH

## A. General formulation of the asymptotic Anderson localization problem

In order to understand the non-universal behaviors found above, we consider a system of $N$ random layers embedded in a background medium with a potential $V = 0$. We will use a scattering element approach, which is best illustrated by an example as shown in Fig. 7(a) which shows three random layers with potentials $v_{i-1}$, $v_i$ and $v_{i+1}$ embedded in a $V = 0$ background. We define a scattering element as part of the sample which starts from the center of one random layer to that of the next random layer. Two such scattering elements are marked by red and blue colors in Fig. 7(a), respectively. Using the standard transfer-matrix method, one can always obtain the transmission and reflection amplitudes across each of the scattering elements. As denoted in Fig. 7(b), we let $t_{1+}$ ($t_{2+}$) and $r_{1+}$ ($r_{2+}$) be the transmission and reflection amplitudes for Element 1 (2) for waves incident from the left (forward waves), and $t_{1-}$ ($t_{2-}$) and $r_{1-}$ ($r_{2-}$) for waves incident from the right (backward

waves). The transmission amplitude $t_{12+}$ through two successive elements (Elements 1 & 2) can be

obtained by summing up all multiply reflected waves between two neighboring scattering elements:

$$t_{12+} = t_{1+}t_{2+} + t_{1+}r_{2+}r_{1-}t_{2+} + t_{1+}r_{2+}r_{1-}r_{2+}r_{1-}t_{2+} + \cdots = \frac{t_{1+}t_{2+}}{1 - r_{2+}r_{1-}} \; . \tag{2}$$

All relevant phases have been included in the reflection and transmission amplitudes. From Eq. (3), we

obtain the transmission $T_{12} = |t_{12+}|^2$ and

$$\ln T_{12} = \ln |t_{12+}|^2 = \ln |t_{1+}|^2 + \ln |t_{2+}|^2 - 2\ln |1 - r_{2+}r_{1-}| \; . \tag{3}$$

Equation (4) also applies recursively to a sample containing $i$ scattering elements. The first $i$-1 random

elements and the $i$-th random element can be treated respectively as Elements 1 and 2. Thus, Eq. (4)

can also represent the transmission of $i$ scattering elements with $t_{1+}$ and $r_{1-}$ replaced by $t^+(i-1)$

and $r^-(i-1)$, respectively, denoting the forward transmission amplitude and backward reflection amplitude of the first $i$-1 scattering elements, and $t_{2+}$ and $r_{2+}$ replaced by $t_{i+}$ and $r_{i+}$, respectively,

denoting the forward transmission and reflection amplitudes of the $i$-th scattering element. By applying

the recursion equation (4) iteratively, we can express the transmission $T_N$ through the system of $N$

random layers as

$$\ln T_N = \sum_{i=1}^{N} \ln |t_{i+}|^2 - 2\sum_{i=2}^{N} \ln |1 - r_{i+}r^-(i-1)| \; . \tag{4}$$

Since the localization length diverges in the long wavelength limit, we should be able to have the reflection amplitude $r_{i\pm} \to 0$ with increasing wavelength through a proper choice of scattering elements. To the leading order of $1/\lambda$, the reflection amplitude can take the form of $r_{i\pm} = C_i^{\pm}\lambda^{-s}$ with

$s > 0$, which allows us to express the Lyapunov exponent in the following form, according to Eqs. (2)

and (5) (see Section 1 of Supplemental Material for details [31]),

$$\gamma \approx \lambda^{-2s} \lim_{N \to \infty} \frac{1}{2L_N} \left[ \sum_{i=1}^{N} \left\langle \left| C_i^+ \right|^2 \right\rangle_c - 2 \sum_{i=2}^{N} \left\langle \Re \left( C_i^+ C_{i-1}^- + \sum_{j=1}^{i-2} C_i^+ C_j^- e^{i\phi_{j+1,i-1}} \right) \right\rangle_c \right], \qquad (5)$$

where $\Re(f)$ denotes the real part of $f$, and $\phi_{j+1,i-1} = \sum_{l=j+1}^{i-1} (\phi_{l+} + \phi_{l-})$ with $\phi_{l\pm}$ being the phase of the transmission amplitude $t_{l\pm}$ of the $l$-th scattering element. Equation (6) gives directly the Anderson localization exponent $m = 2s$ $(\xi = \gamma^{-1} \propto \lambda^m)$. For pseudospin-1 systems, the choice of scattering elements is shown in Fig. 7(a). It will be shown later that the value of $s$ depends on the type of random potential. For other systems, we may take a different choice of scattering elements to have the reflection amplitude in the form of $r_{i\pm} = C_i^\pm \lambda^{-s}$. It should be pointed out that the above choice of scattering elements makes it very convenient to obtain the localization length exponent through Eq. (6) as the bracket on its right-hand side does not contribute to the exponent. If other scattering elements are chosen which do not vanish in the long wavelength limit, Eq. (5), instead of Eq. (6) has to be used to numerically calculate the Anderson localization length behavior, although the same result is expected.

## B. Asymptotic localization length for ordinary 1D disordered systems

To demonstrate the usefulness of our method and to illustrate how it works, we use it to derive the well-known long wavelength behavior of an ordinary 1D disordered system which consists of $N$ randomly distributed dielectric layers with the same thickness $d$. The relative permittivity $\varepsilon_i$ in the $i$-th random layer is a random number fluctuating around the permittivity $\varepsilon_b$ of the background medium, i.e., $\left\langle \Delta \varepsilon_i \right\rangle_c = \left\langle \varepsilon_i - \varepsilon_b \right\rangle_c = 0$. To use Eq. (6) properly, it is important to choose a scattering element so that its reflection amplitude vanishes in the long wavelength limit ($\lambda \to \infty$). For example, in the case of layered random dielectric systems studied here, if we take a scattering element from the center of one layer to that of next layer, as shown in Fig. 7(c), the reflection amplitude is mainly determined by the impedance mismatch between the two layers, and it is non-zero, independent of the wavelength. To have the reflection amplitude in the form of $r_{i\pm} = C_i^\pm \lambda^{-s}$ at long wavelengths, we insert a background layer with arbitrarily small thickness $\delta d \to 0$ in between any two neighboring random dielectric layers and take the $i$-th scattering element as the $i$-th random layer embedded in the background medium of vanishing thickness, as shown in Fig. 7(d). Note that as shown in section 2.1 of Supplementary Material [31], inserting a background layer with zero thickness makes no difference on the total transfer matrix of the system. The transfer matrix for the $i$-th scattering element at normal incidence ($\theta = 0$) has the form

$$M_i = \begin{pmatrix} \cos k_i d + \dfrac{i}{2}(z_i + \dfrac{1}{z_i})\sin k_i d & -\dfrac{i}{2}(z_i - \dfrac{1}{z_i})\sin k_i d \\ \dfrac{i}{2}(z_i - \dfrac{1}{z_i})\sin k_i d & \cos k_i d - \dfrac{i}{2}(z_i + \dfrac{1}{z_i})\sin k_i d \end{pmatrix}, \qquad (6)$$

where $k_i = \sqrt{\varepsilon_i}\,\omega/c$ is the wavenumber in the $i$-th dielectric layer, $z_i = \sqrt{\varepsilon_b/\varepsilon_i}$ is the relative impedance of the $i$-th dielectric layer with respect to the background medium, $\omega$ is the angular frequency and $c$ is the speed of light in vacuum. At long wavelengths, Eq. (7) reduces to

$$M_i(\lambda \to \infty) \approx \begin{pmatrix} 1 & i\dfrac{\pi d}{\lambda}\dfrac{\Delta\varepsilon_i}{\sqrt{\varepsilon_b}} \\ -i\dfrac{\pi d}{\lambda}\dfrac{\Delta\varepsilon_i}{\sqrt{\varepsilon_b}} & 1 \end{pmatrix}. \qquad (7)$$

Thus, the above choice of scattering elements ensures that the reflection amplitude obtained from $M_i$ indeed follows the form, $r_{i\pm} = C_i^{\pm}\lambda^{-s}$ when $\omega \to 0$ or $\lambda \to \infty$ ($\lambda = 2\pi c/\omega$), i.e.,

$r_{i\pm} = -\dfrac{(M_i)_{21}}{(M_i)_{22}} \approx i\pi\dfrac{\Delta\varepsilon_i d}{\sqrt{\varepsilon_b}}\lambda^{-1}$. The transmission amplitude can also be obtained from Eq. (8) as $t_{i\pm} \approx 1$ when $\lambda \to \infty$, i.e., the transmission phase is $\phi_{t\pm} \approx 0$. By substituting $s = 1$, $C_i^{\pm} = i\pi\dfrac{\Delta\varepsilon_i d}{\sqrt{\varepsilon_b}}$ and

$\phi_{t\pm} \approx 0$ into Eq. (6), we obtain $\gamma = \dfrac{\pi^2 d}{2\varepsilon_b}\left\langle(\Delta\varepsilon_i)^2\right\rangle_c \lambda^{-2}$ or $\xi = \dfrac{2\varepsilon_b}{\pi^2 d}\dfrac{\lambda^2}{\left\langle(\Delta\varepsilon_i)^2\right\rangle_c}$ for any type of uncorrelated disorder, i.e., $<\Delta\varepsilon_i\Delta\varepsilon_j>_c = \delta_{i,j}<(\Delta\varepsilon_i)^2>_c$, which agrees with the results for ordinary 1D disordered systems in literature [5-10].

## C. Asymptotic localization length for pseudospin-1 systems

For pseudospin-1 systems subjected to 1D random potentials, we choose a scattering element from the center of one random layer to that of the next random layer, as shown by the colored region in Fig. 7(a). Next, we use the TMM to obtain the reflection amplitudes and transmission phases of the $i$-th scattering element for normalized incident energy $\bar{E}$ and incident angle $\theta$. Using the following matrix representation of spin-1 operator, $\vec{S} = S_x \hat{x} + S_y \hat{y}$,

$$S_x = \frac{1}{\sqrt{2}} \begin{pmatrix} 0 & 1 & 0 \\ 1 & 0 & 1 \\ 0 & 1 & 0 \end{pmatrix}, \qquad S_y = \frac{1}{\sqrt{2}} \begin{pmatrix} 0 & -i & 0 \\ i & 0 & -i \\ 0 & i & 0 \end{pmatrix}, \tag{8}$$

we can give the eigenvectors of the pseudospin-1 Hamiltonian $H = \hbar v_g \vec{S} \cdot \vec{k} + V$ for a homogeneous medium with a constant potential $V$ as [41, 51]

$$\psi_{s',\vec{k}}(\vec{r}) = \frac{1}{2} \begin{pmatrix} s' e^{-i\theta_k} \\ \sqrt{2} \\ s' e^{i\theta_k} \end{pmatrix} e^{i\vec{k} \cdot \vec{r}}, \tag{9}$$

where $s' = \mathrm{sgn}(\bar{E} - \bar{V})$ and $\theta_k$ is the angle of the wavevector $\vec{k}$ with respect to the $x$-axis. Note that here $\bar{E} = E / \hbar v_g$ and $\bar{V} = V / \hbar v_g$ are normalized incident energy and potential, respectively. In the case of 1D layered structure, the wavevector component parallel to the interface $k_y = k_b \sin\theta$ is a conserved quantity, where $k_b = \bar{E}$ is the wavevector in the background medium. Thus, the wave function $\psi = (\psi_1, \psi_2, \psi_3)^T$ in different regions of Fig. 7(a) can be written in terms of the incident and reflected waves using the eigenvectors in Eq. (10) as follows. In region I, we have

$$\psi_1 = \frac{\tilde{a}_{i-1}}{2} \begin{pmatrix} s_{i-1} e^{-i\tilde{\theta}_{i-1}} \\ \sqrt{2} \\ s_{i-1} e^{i\tilde{\theta}_{i-1}} \end{pmatrix} e^{i(k_{i-1,x} x + k_y y)} + \frac{\tilde{b}_{i-1}}{2} \begin{pmatrix} s_{i-1} e^{-i(\pi - \tilde{\theta}_{i-1})} \\ \sqrt{2} \\ s_{i-1} e^{i(\pi - \tilde{\theta}_{i-1})} \end{pmatrix} e^{i(-k_{i-1,x} x + k_y y)}, \tag{10}$$

where $\tilde{\theta}_{i-1}$ is the angle of the incident wavevector $\vec{k}_{i-1}$ with respect to the $x$-axis in the $(i\text{-}1)$-th random layer, $s_{i-1} = \text{sgn}(\bar{E} - \bar{v}_{i-1})$ and $k_{i-1,x}$ is the $x$ component of $\vec{k}_{i-1}$ with

$k_{i-1,x} = s_{i-1}\sqrt{(\bar{E} - \bar{v}_{i-1})^2 - k_y^2} = |\bar{E} - \bar{v}_{i-1}|\cos\tilde{\theta}_{i-1}$. In region II, we have

$$\psi_{\text{II}} = \frac{\tilde{a}_b}{2}\begin{pmatrix} s_b e^{-i\tilde{\theta}_b} \\ \sqrt{2} \\ s_b e^{i\tilde{\theta}_b} \end{pmatrix} e^{i(k_{bx}x + k_y y)} + \frac{\tilde{b}_b}{2}\begin{pmatrix} s_b e^{-i(\pi - \tilde{\theta}_b)} \\ \sqrt{2} \\ s_b e^{i(\pi - \tilde{\theta}_b)} \end{pmatrix} e^{i(-k_{bx}x + k_y y)}, \tag{11}$$

where $\tilde{\theta}_b$ is the angle of the incident wavevector $\vec{k}_b$ with respect to the $x$-axis in the background layer, $s_b = \text{sgn}(\bar{E})$ and $k_{bx}$ is the $x$ component of $\vec{k}_b$ with $k_{bx} = s_b\sqrt{\bar{E}^2 - k_y^2} = |\bar{E}|\cos\tilde{\theta}_b$. Note that the incident angle $\theta$ refers to the direction of incident energy/current flux in the background medium,

which can be either parallel or antiparallel to the incident wavevector $\vec{k}_b$, depending on the sign of incident energy $s_b = \text{sgn}(\bar{E})$, i.e., whether the band is a positive band or a negative band. In region III,

we have

$$\psi_{\text{III}} = \frac{\tilde{a}_i}{2}\begin{pmatrix} s_i e^{-i\tilde{\theta}_i} \\ \sqrt{2} \\ s_i e^{i\tilde{\theta}_i} \end{pmatrix} e^{i(k_{i,x}x + k_y y)} + \frac{\tilde{b}_i}{2}\begin{pmatrix} s_i e^{-i(\pi - \tilde{\theta}_i)} \\ \sqrt{2} \\ s_i e^{i(\pi - \tilde{\theta}_i)} \end{pmatrix} e^{i(-k_{i,x}x + k_y y)}. \tag{12}$$

We define the transfer matrix $M^{(1)}(i)$ of the $i$-th scattering element by the relation,

$$\begin{pmatrix} a_i \\ b_i \end{pmatrix} = M^{(1)}(i)\begin{pmatrix} a_{i-1} \\ b_{i-1} \end{pmatrix}, \tag{13}$$

with $a_{i-1} = \tilde{a}_{i-1}e^{ik_{i-1,x}x_0}$, $b_{i-1} = \tilde{b}_{i-1}e^{-ik_{i-1,x}x_0}$, $a_i = \tilde{a}_i e^{ik_{i,x}(x_0 + d + \Delta)}$ and $b_i = \tilde{b}_i e^{-ik_{i,x}(x_0 + d + \Delta)}$. Here $x = x_0$ and

$x = x_0 + d + \Delta$ are the $x$ coordinates of the left and right ends of the $i$-th scattering element,

respectively, as shown in Fig. 7(a). Applying the boundary conditions of pseudospin-1 systems that

$\psi_2$ and $\psi_1 + \psi_3$ are continuous at the boundaries between two neighboring regions, we can obtain

the transfer matrix $M^{(1)}(i)$,

$$M^{(1)}(i) = \begin{pmatrix} e^{i[\Phi(i-1)+\Phi(i)]}\left(\alpha_{i+}^{(1)} + i\beta_{i+}^{(1)}\right)\Lambda_i^{-1} & e^{i[-\Phi(i-1)+\Phi(i)]}\left(\alpha_{i-}^{(1)} - i\beta_{i-}^{(1)}\right)\Lambda_i^{-1} \\ e^{i[\Phi(i-1)-\Phi(i)]}\left(\alpha_{i-}^{(1)} + i\beta_{i-}^{(1)}\right)\Lambda_i^{-1} & e^{-i[\Phi(i-1)+\Phi(i)]}\left(\alpha_{i+}^{(1)} - i\beta_{i+}^{(1)}\right)\Lambda_i^{-1} \end{pmatrix}, \tag{14}$$

with $\Phi(i)=\dfrac{d}{2}\left(\bar{E}-\bar{v}_i\right)\cos\theta_i$, $\alpha_{i\pm}^{(1)}=\left(\cos\theta_i\pm\cos\theta_{i-1}\right)\cos\theta\cos\left(\bar{E}\Delta\cos\theta\right)$,

$\beta_{i\pm}^{(1)}=\left(\cos\theta_{i-1}\cos\theta_i\pm\cos^2\theta\right)\sin\left(\bar{E}\Delta\cos\theta\right)$, $\Lambda_i=2\cos\theta_i\cos\theta$ and

$\cos\theta_i=s_i\cos\tilde{\theta}_i=\sqrt{1-\dfrac{\bar{E}^2\sin^2\theta}{(\bar{E}-\bar{v}_i)^2}}$. The reflection and transmission amplitudes can be obtained from the

transfer matrix $M^{(1)}(i)$ as

$$r_{i+}=-e^{2i\Phi(i-1)}\frac{\alpha_{i-}^{(1)}+i\beta_{i-}^{(1)}}{\alpha_{i+}^{(1)}-i\beta_{i+}^{(1)}}\,,\tag{15}$$

$$r_{i-}=e^{2i\Phi(i)}\frac{\alpha_{i-}^{(1)}-i\beta_{i-}^{(1)}}{\alpha_{i+}^{(1)}-i\beta_{i+}^{(1)}}\,,\tag{16}$$

$$t_{i+}=\frac{\Lambda_{i-1}}{\alpha_{i+}^{(1)}-i\beta_{i+}^{(1)}}e^{i[\Phi(i-1)+\Phi(i)]}\,,\tag{17}$$

$$t_{i-}=\frac{\Lambda_i}{\alpha_{i+}^{(1)}-i\beta_{i+}^{(1)}}e^{i[\Phi(i-1)+\Phi(i)]}\,.\tag{18}$$

As can be seen in Figs. 1(b)-(d), in the case of Types I and II disorder, the width of the background

layer in the chosen scattering element shown in Fig. 7(a) is $\Delta=0$, whereas in the case of Type III

disorder, $\Delta=d$. Note that $\beta_{i\pm}^{(1)}=0$ when $\Delta=0$. Thus, for the types of disorder with $\Delta=0$, e.g.,

Types I and II disorder, we can simplify the reflection amplitudes and transmission phases as

$$r_{i+}(\Delta=0)=-e^{id(\bar{E}-\bar{v}_{i-1})\cos\theta_i}\frac{\cos\theta_i-\cos\theta_{i-1}}{\cos\theta_i+\cos\theta_{i-1}}\,,\quad,\tag{19}$$

$$r_{i-}(\Delta=0)=e^{id(\bar{E}-\bar{v}_i)\cos\theta_i}\frac{\cos\theta_i-\cos\theta_{i-1}}{\cos\theta_i+\cos\theta_{i-1}}\,,\tag{20}$$

$$\phi_{i\pm}(\Delta=0)=\frac{d}{2}\left((\bar{E}-\bar{v}_{i-1})\cos\theta_{i-1}+(\bar{E}-\bar{v}_i)\cos\theta_i\right)\,,\tag{21}$$

with $\cos\theta_i=\sqrt{1-\dfrac{\bar{E}^2\sin^2\theta}{(\bar{E}-\bar{v}_i)^2}}$. For biased binary disorder, where $\bar{v}_i$ is either $\bar{V}_0+\bar{W}$ or $\bar{V}_0-\bar{W}$, we have $\cos\theta_i=\cos\theta_{i-1}$ at $\bar{E}=\bar{V}_0$, regardless of the value of $\theta$, which in turn leads to zero reflections

for all incident angles, as shown in Eqs. (20) and (21). This is a direct manifestation of the SKTE in

pseudospin-1 systems. In the long wavelength limit ($\bar{E}\to0$), $\cos\theta_i$ can be approximated as

$\cos\theta_i\approx1-\dfrac{\bar{E}^2\sin^2\theta}{2(\bar{E}-\bar{v}_i)^2}$. Thus, the reflection amplitudes in Eqs. (20) and (21) can be expressed as

$$r_{i+}(\Delta=0)\approx\frac{\bar{E}^2\sin^2\theta}{4}e^{id(\bar{E}-\bar{v}_{i-1})\cos\theta_{i-1}}\left[\frac{1}{(\bar{E}-\bar{v}_i)^2}-\frac{1}{(\bar{E}-\bar{v}_{i-1})^2}\right]\,,\tag{22}$$

$$r_{i-}(\Delta = 0) \approx -\frac{\bar{E}^2 \sin^2 \theta}{4} e^{id(\bar{E}-\bar{v}_i)\cos\theta_i} \left[ \frac{1}{(\bar{E}-\bar{v}_i)^2} - \frac{1}{(\bar{E}-\bar{v}_{i-1})^2} \right]. \tag{23}$$

It is easy to show that as $\bar{E} \to 0$, Eqs. (23) and (24) give $r_{i\pm} \propto \bar{E}^3$ for binary disorder and $r_{i\pm} \propto \bar{E}^2$ for biased binary disorder, which explain the $\xi \propto \lambda^6$ and $\xi \propto \lambda^4$ behaviors found in Figs. 2(a) and 2(c), respectively, through Eq. (6). According to Eqs. (22)-(24), the exact asymptotic solution of $\gamma$ can be obtained by evaluating the expression in the bracket of Eq. (6). For biased binary disorder, we find (see section 2.2.1 of Supplemental Material for details [31])

$$\gamma = \xi^{-1} \approx \frac{4p(1-p)\bar{E}^4 \sin^4 \theta}{d} \frac{\bar{W}^2(\bar{E}-\bar{V}_0)^2}{\left[ (\bar{E}-\bar{V}_0)^2 - \bar{W}^2 \right]^4} F(\varphi_+, \varphi_-, s_+, s_-), \tag{24}$$

where

$$F(\varphi_+, \varphi_-, s_+, s_-) = \frac{\sin^2 \varphi_+ \sin^2 \varphi_-}{1 - 2p(1-p)\sin^2(s_+\varphi_+ - s_-\varphi_-) - p\cos 2\varphi_+ - (1-p)\cos 2\varphi_-}. \tag{25}$$

Here $\varphi_{\pm} = \sqrt{(\bar{E}-\bar{V}_0 \mp \bar{W})^2 - \bar{E}^2 \sin^2 \theta} d$ and $s_{\pm} = \text{sgn}(\bar{E}-\bar{V}_0 \mp \bar{W})$. Results of Eq. (25) for $p = 0.5$ are shown by solid curves in Figs. 2(a) and 2(c) for the cases of $\bar{V}_0 = 0$ and $\bar{V}_0 = 0.1\pi / d$, respectively. Excellent quantitative agreements are found between the analytical and numerical results for a wide range of wavelength, not limited to the long-wavelength regime. For $\bar{V}_0 \neq 0$, since Eq. (25) contains both the factors, $\bar{E}^4$ and $(\bar{E}-\bar{V}_0)^2$, it gives explicitly the asymptotic behaviors $\xi \propto \lambda^4$ and $\xi \propto (\bar{E}-\bar{V}_0)^{-2}$ found numerically in Figs. 2(c) and (3). The divergence at $\bar{E} = \bar{V}_0 \neq 0$ is due to the SKTE, which merges with the former factor when $\bar{V}_0 = 0$, leading to $\xi \propto \lambda^6$ divergence in Fig. 2(a) for Type I disorder.

For Type II disorder, the SKTE breaks down due to the presence of additional randomness, and Eqs. (23) and (24) give $r_{i\pm} \propto \bar{E}^2$ when $\bar{E} \to 0$. According to Eqs. (22)-(24), the long-wavelength analytical expression of $\gamma$ can be obtained through Eq. (6). For Type II disorder, we find (see section

2.2.2 of Supplemental Material for mathematical details [31]),

$$\gamma = \frac{Q^2 \sin^4 \theta}{12d} \left(1 - \cos 2\bar{W}d\right) \left(\frac{\bar{E}}{\bar{W}}\right)^4 + \frac{2p(1-p)\sin^4 \theta \sin^2 \bar{W}d}{\left[1 - 4p(1-p)\cos^2 \bar{W}d\right]d} \left(\frac{\bar{E}}{\bar{W}}\right)^6. \qquad (25)$$

Note that when $\bar{E}/\bar{W} << Q$, the first term in Eq. (27) dominates, which gives the $\xi \propto \lambda^4$ behavior in

Fig. 2(b). However, when $\bar{E}/\bar{W} >> Q$, the higher order term $\propto \left(\bar{E}/\bar{W}\right)^6$ dominates. Thus, $\xi$

crosses over from $\lambda^6$ when $\lambda << \frac{2\pi}{Q\bar{W}}$ to $\lambda^4$ when $\lambda$ is increased to $\lambda >> \frac{2\pi}{Q\bar{W}}$, as shown in

Fig. 2(b).

For Type III disorder, the width of the background layer is $\Delta = d \neq 0$ and the normalized random potential is $\bar{v}_i = \bar{U}_0(1 + \delta_i)$. As can be seen from Eqs. (16) and (17), the multiple reflections inside the background layer introduce an additional term $\beta_{i-}^{(1)} = \left(\cos\theta_{i-1} \cos\theta_i - \cos^2\theta\right)\sin\left(\bar{E}\Delta\cos\theta\right)$ in the numerators of $r_{i\pm}$, which is linear in $\bar{E}$ when $\bar{E} \to 0$. Thus, the reflection amplitudes and transmission phases now become $r_{i+} \approx -\frac{i\bar{E}}{2}e^{-i\bar{v}_{i-1}d}\Delta\sin^2\theta$, $r_{i-} \approx -\frac{i\bar{E}}{2}e^{-i\bar{v}_i d}\Delta\sin^2\theta$ and

$\phi_{i\pm} = -d(\bar{v}_{i-1} + \bar{v}_i)/2$ in the long wavelength limit. The analytical expression of $\gamma$ can be obtained

using Eq. (6) (see section 2.2.3 of Supplemental Material for mathematical details [31]):

$$\gamma \approx \bar{E}^2 \frac{\bar{U}_0^2 d^3 Q^2 \sin^4 \theta}{48 \sin^2 \bar{U}_0 d}, \qquad (26)$$

showing exactly the same behaviors in Fig. 2(d). The $\xi \propto \lambda^2$ behavior actually holds for any type of random potential, as long as the random layers are separated by layers of background medium (see section 2.2.3 of Supplemental Material [31]). It should be pointed out that for the case of uniform disorder, for which the normalized potential $\bar{v}$ in each layer is an independent random number distributed uniformly on the interval [ $-\bar{W}$, $\bar{W}$ ], the long-wavelength localization length also diverges as $\lambda^2$ [24]. However, its underlying origin is the emergence of evanescent waves in the system [24], not the Anderson localization for Type III disorder considered here.

For pseudospin-1/2 systems subjected to 1D random potentials, we take the same choice of scattering elements as that for pseudospin-1 systems, as shown in Fig. 7(a). Similarly, we can employ the Pauli matrices for the spin-1/2 operator to obtain the transfer matrix of individual scattering elements and derive the exact asymptotic expressions of the Lyapunov exponent for Types I-III disorder (see section 2.3 of Supplemental Material [31]). Due to different set of boundary conditions, there is no SKTE for pseudospin-1/2 systems and the reflection amplitude follows $r_{\pm} \propto \bar{E}$ when $\bar{E} \to 0$ for all types of disorder, which gives $\xi \propto \lambda^2$ behaviors through Eq. (6) (see section 2.3 of Supplemental Material [31]). The analytical results are plotted as solid curves in Fig. 5 to compare with the numerical results calculated by the TMM. Again, our analytical asymptotic solutions agree quantitatively with numerical results.

# V. CONCLUSIONS

We discovered non-universal critical behaviors for pseudospin-1 systems when the system is subjected to different types of disorder. In contrast, the critical exponents for pseudospin-1/2 systems always follow $\xi \propto \lambda^2$ in the long wavelength limit for all types of disorder, same as ordinary 1D disordered systems. The critical exponents can be predicted using an analytical method based on a stack recursion equation and the phenomena can be interpreted using the scattering properties of the scattering elements. Our new analytical method proposed here provides a simple way to determine the critical exponent *m* through the long-wavelength scattering properties of properly chosen scattering elements. It is applicable to general 1D Anderson localization problems as long as the localization length diverges at long wavelengths. The transport mean free path is also calculated using the TMM for both pseudospin systems. It is found that the transport mean free path and the Anderson localization length are of the same order of magnitude for both systems, indicating the absence of a diffusive regime, same as the known results in ordinary 1D disordered systems.

# VI. ACKNOWLEDGEMENTS

This work was supported by a grant from the Research Grants Council of Hong Kong (Project

AoE/P-02/12). S. G. L. also acknowledges support by the National Science Foundation under grant

DMR-1508412.

# Figure Captions

Fig. 1 (a) Geometry of the disordered structure. (b)-(d) Schematics of Types I-III disorder (see text).

Fig. 2 (Color online) Localization length as a function of wavelength at two different incident angles for pseudospin-1 systems subjected to different types of disorder (see text): (a) Type I disorder with $p = 0.5$ and $\bar{W} = 0.2\pi/d$; (b) Type II disorder with $p = 0.5$, $\bar{W} = 0.2\pi/d$ and $Q = 0.3$; (c) biased binary disorder with $p = 0.5$, $\bar{W} = 0.2\pi/d$ and $\bar{V}_0 = 0.1\pi/d$; and (d) Type III disorder with $Q = 0.3$ and $\bar{U}_0 = 0.2\pi/d$. The symbols are numerical results calculated using the TMM and red solid lines are analytical results obtained by using Eq. (6).

Fig. 3 (Color online) Localization length as a function of $|\bar{E} - \bar{V}_0|$ for two incident angles for biased binary disorder with $p = 0.5$, $\bar{W} = 0.2\pi/d$ and $\bar{V}_0 = 0.1\pi/d$. The symbols are numerical results calculated using the TMM, which are fitted by $\xi \propto |\bar{E} - \bar{V}_0|^{-2}$ (red dotted lines) as $\bar{E} \to \bar{V}_0$.

Fig. 4 (Color online) Transport mean free path as a function of wavelength at two different incident angles for pseudospin-1 systems subjected to different types of disorder (see text): (a) Type I disorder with $p = 0.5$ and $\bar{W} = 0.2\pi/d$; (b) Type II disorder with $p = 0.5$, $\bar{W} = 0.2\pi/d$ and $Q = 0.3$; (c) Type III disorder with $Q = 0.3$ and $\bar{U}_0 = 0.2\pi/d$. The symbols are numerical results calculated using the TMM, which are well fitted by red dotted lines at long wavelengths, showing $l_t \propto \lambda^m$ with $m = 6$, 4 and 2 for Types I-III disorder, respectively.

Fig. 5 (Color online) Same as Fig. 2, but for pseudospin-1/2 systems.

Fig. 6 (Color online) Same as Fig. 4, but for pseudospin-1/2 systems. All transport mean free paths are well fitted by $l_t \propto \lambda^2$ (red dotted lines) at long wavelengths.

Fig. 7 (Color online) (a) Schematic of the choice of scattering elements in pseudospin-1 systems. (b) Schematic of the transmissions and reflections for Element 1, Element 2 and their combination. (c) Schematic of the choice of scattering elements from the center of one random layer to that of next layer for ordinary 1D disordered systems. (d) Schematic of the choice of scattering elements with one random dielectric layer embedded in the background medium of vanishing thickness. Note that the inserted background layer in between two random dielectric layers has a vanishing thickness $\delta d = 0$.

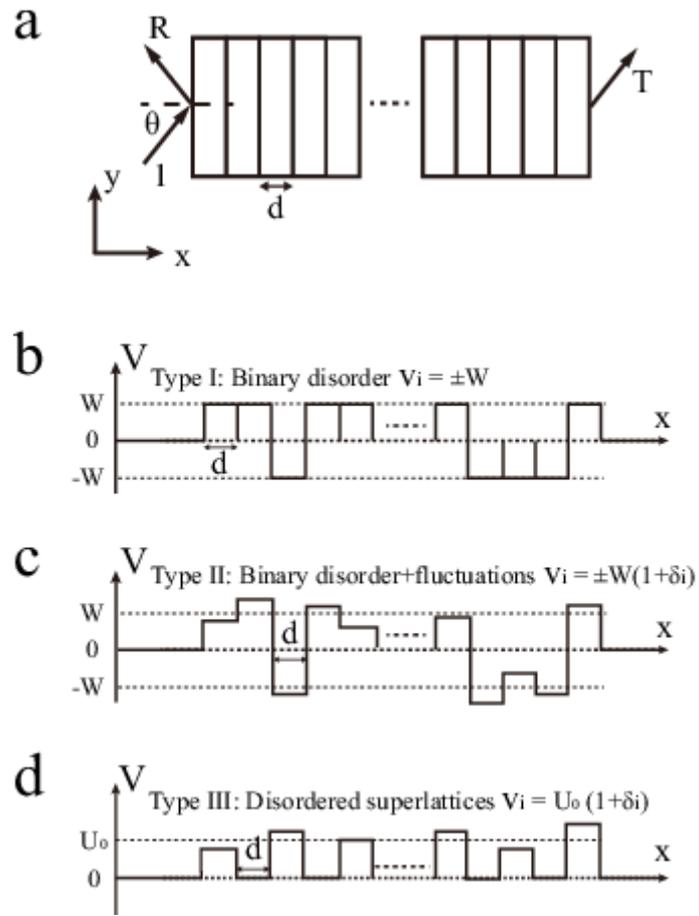

Fig. 1

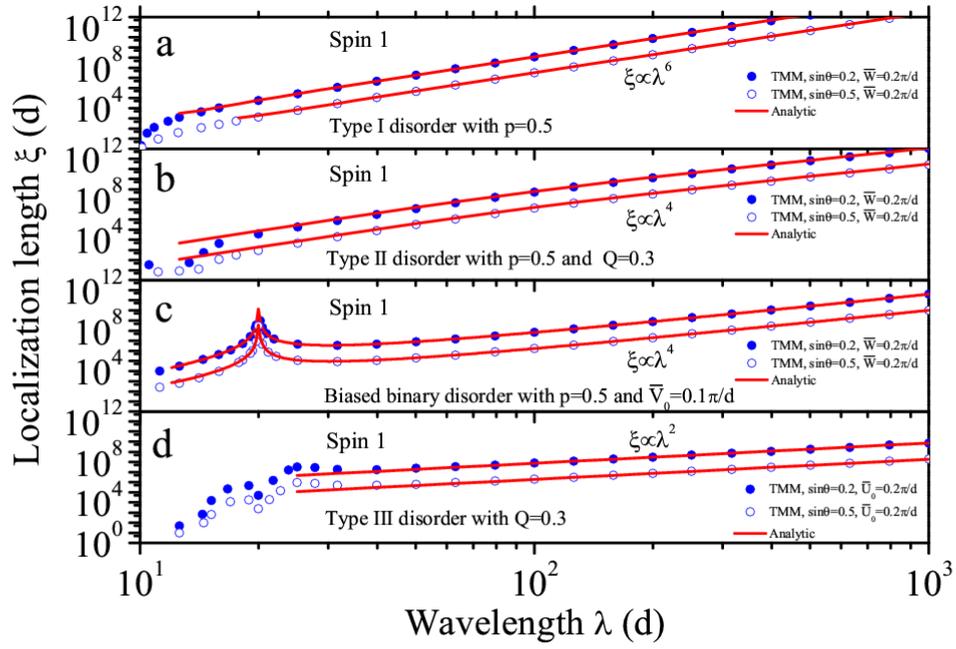

Fig. 2

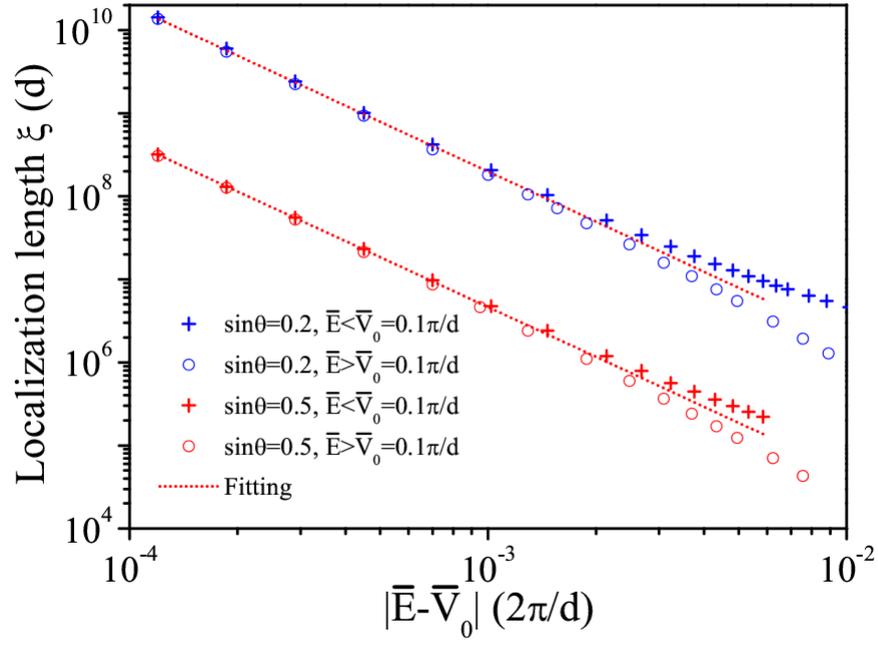

Fig. 3

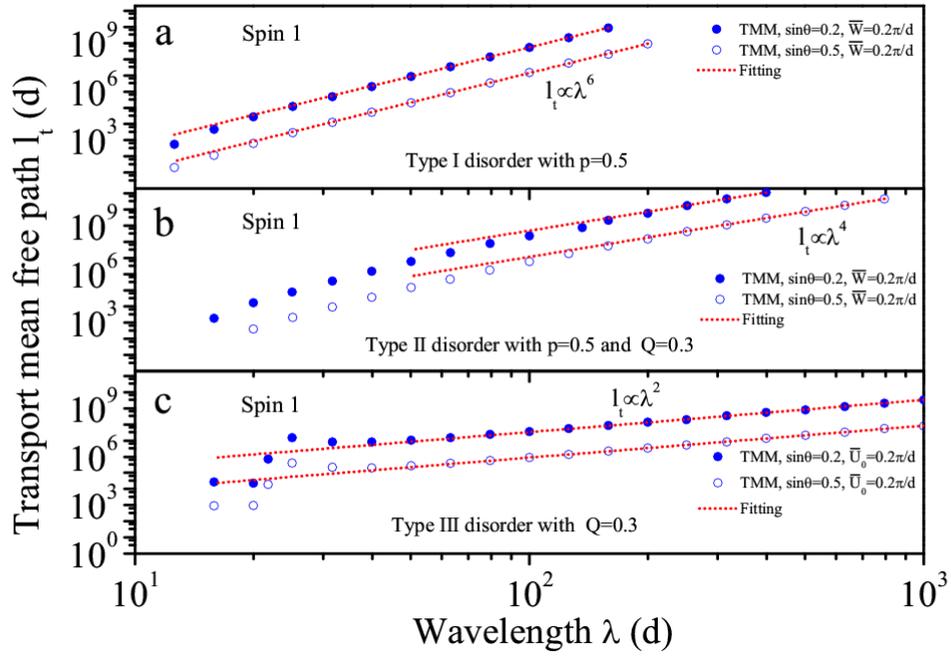

Fig. 4

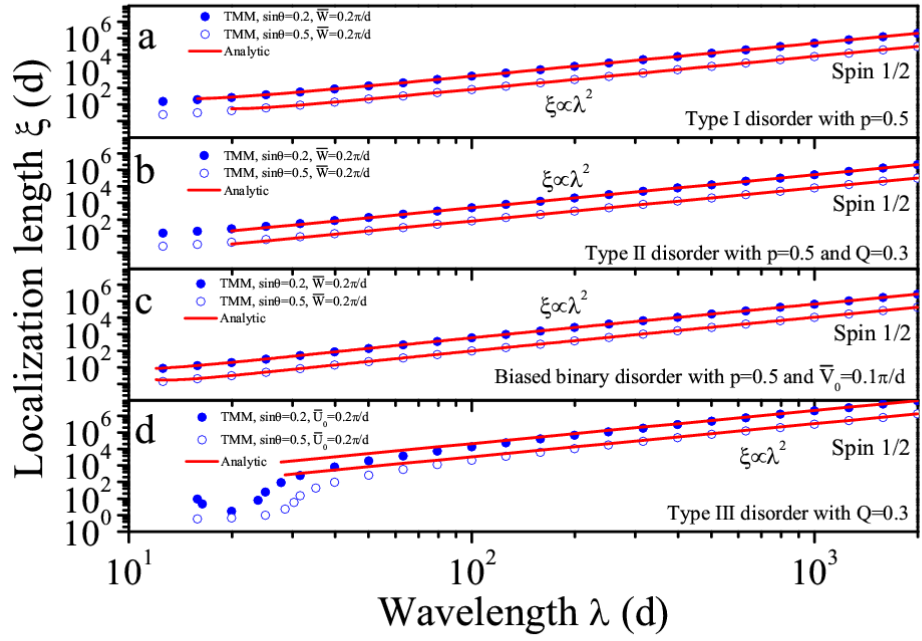

Fig. 5

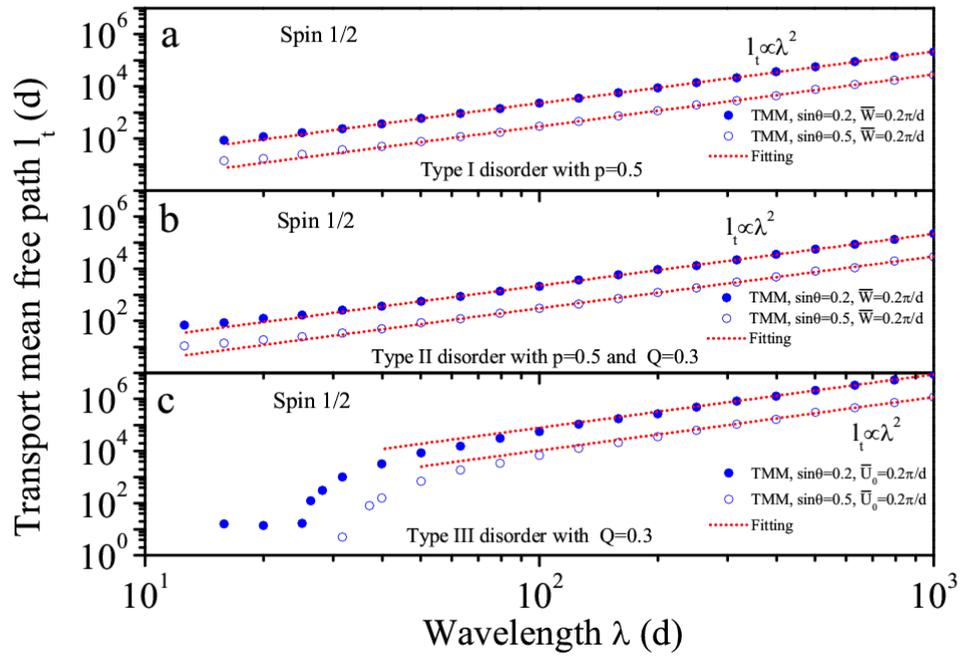

Fig. 6

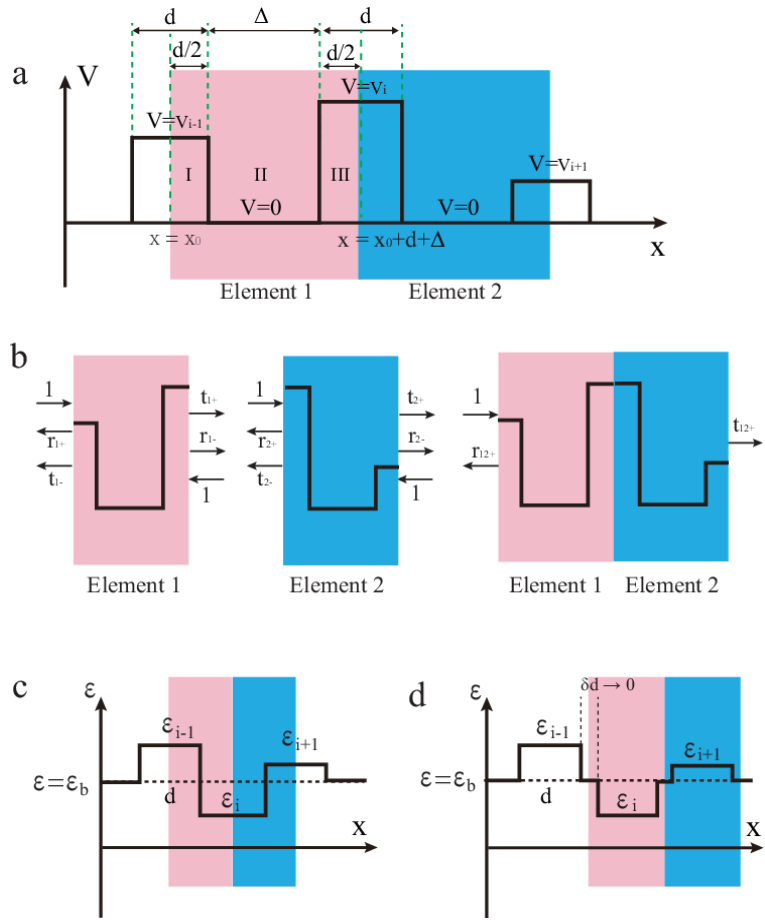

Fig. 7